\documentclass[preprint,showpacs,preprintnumbers,amsmath,amssymb]{revtex4}
\usepackage{graphicx}
\begin{document}
\title{Magnetic susceptibility, exchange interactions and spin-wave spectra in the
local spin density approximation}
\author{M I Katsnelson$^{1,2}$ and A I Lichtenstein$^{2,3}$}
\address{$^{1}$ Department of Physics, Uppsala University, Box 530, SE-75121 Uppsala,
Sweden}
\address{$^{2}$Department of Physics, University of Nijmegen,
NL-6525 ED Nijmegen, The Netherlands}
\address{$^{3}$ Institute of Theoretical Physics, University of Hamburg,
Jungiusstrasse 9, 20355 Hamburg, Germany}
\date{\today}

\begin{abstract}
Starting from exact expression for the dynamical spin
susceptibility in the time-dependent density functional theory a
controversial issue about exchange interaction parameters and
spin-wave excitation spectra of itinerant electron ferromagnets is
reconsidered. It is shown that the original expressions for
exchange integrals based on the magnetic force theorem (J. Phys.
F {\bf 14} L125 (1984)) are optimal for the calculations of the
magnon spectrum whereas static response function is better
described by the ``renormalized'' magnetic force theorem by P.
Bruno (Phys. Rev. Lett. {\bf 90} 087205 (2003)). This conclusion
is confirmed by the {\it ab initio} calculations for Fe and Ni.
\end{abstract}


\maketitle

An efficient scheme for the first-principle calculations of
exchange interaction parameters in magnets based on a so-called
``magnetic force theorem'' (MFT) \cite{LIXT,JMMM} in the density
functional theory is frequently used for analysis of exchange
parameters for different classes of magnetic materials such as
dilute magnetic semiconductors \cite{DMS}, molecular magnets
\cite{molmag}, colossal magnetoresistance perovskites \cite{CMR},
transition metal alloys \cite{alloy}, hard magnetic materials
such as PtCo \cite{PtCo} and many others. Recently this method
was generalized to take into account the correlation effects and
successfully used for the quantitative estimation of exchange
interactions in Fe and Ni \cite{KL2000,KL2002}. At the same time
the formal status of this approach is still not well-defined
since a general mapping of formally rigorous spin density
functional to an effective classical Heisenberg Hamiltonian can
be done only approximately. It was noticed already in the first
work on the MFT \cite{LIXT} that only the expression for spin
wave stiffness constant $D$ is reliable. In terms of the
diagrammatic many-body approach it means that the exchange
integrals ($J_{ij}$) in general should contain vertex corrections
\cite{Ferdi} which are neglected in the simple expression
\cite{LIXT,JMMM} (see Ref. \onlinecite{KL2002}). At the same
time, using a general expression for spin-wave stiffness due to
Hertz and Edwards \cite{edwards} one can prove that the vertex
corrections to $D$ are cancelled for any local approximation for
the self-energy (or, in the density functional method, for the
local exchange correlation potential) \cite{berlin}. Recently P.
Bruno has suggested \cite{bruno} corrections to the MFT and
consequently to the expressions for $J_{ij}$ (see also Ref.
\onlinecite{antr}). It is important to note that, first, the new
expression for $D$ coincides with the old one and, second, that
these corrections for the case of itinerant electron magnets are
formally small in adiabatic parameter $\eta =\varpi /\Delta$
where $\varpi$ is a characteristic magnon frequency and $\Delta$
is the Stoner spin splitting. At the same time, mapping of the
local spin density approximation (LSDA) onto the classical
Heisenberg model itself is valid only in the adiabatic
approximation $\eta \rightarrow 0$ \cite{LIXT,JMMM,SD}. If we are
interested in higher order effects in the $\eta$ it might be
needed different effective exchange parameters for different
physical properties. We will show on few examples that this is
exactly the case. It turns out that the spin-wave excitation
spectrum should be calculated in terms of ``old'' exchange
integrals \cite{LIXT,JMMM} whereas for static properties ``new''
exchange integrals \cite{bruno,antr} are more appropriate.

The most reliable way to consider spin-wave properties of
itinerant electron magnets in the framework of the spin density
functional theory is the use of frequency dependent magnetic
susceptibility \cite{callaway,cooke,savr}. One should start from
the time-dependent density functional theory in the adiabatic
approximation (ADA-TDDFT) \cite{TDSDF,vosko}. We proceed with the
Schr\"{o}dinger-like equation within the self-consistent ADA-TDDFT
potential
\begin{eqnarray}
i\frac{\partial \psi }{\partial t} &=&H\psi  \nonumber \\
H &=&-\nabla ^{2}+V({\bf r})-\frac{1}{2}({\bf B}_{xc}({\bf r})+{\bf B}_{ext}(%
{\bf r})){\bf \sigma }  \label{schroed}
\end{eqnarray}
(Slater's units are used here) where $V({\bf r})$ is an effective
potential, ${\bf B}_{ext}({\bf r})$ and ${\bf B}_{xc}({\bf r})$
are external magnetic field acting on spin and
exchange-correlation field, respectively. The adiabatic approximation means
that the functional dependences of exchange-correlation potential and field
on the charge and spin density are supposed to be the same as in the stationary
case. In LSDA one has
\begin{eqnarray}
V({\bf r}) &=&V_{ext}({\bf r})+\int d{\bf r}^{\prime }\frac{n({\bf r}%
^{\prime })}{\left| {\bf r-r}^{\prime }\right| }+\frac{\partial }{\partial n}%
[n\varepsilon _{xc}]  \nonumber \\
{\bf B}_{xc} &=&-2\frac{{\bf m}}{m}\frac{\partial }{\partial m}[n\varepsilon
_{xc}]  \label{local}
\end{eqnarray}
where $n$ and ${\bf m}${\bf \ }are charge and spin density,
$\varepsilon _{xc}$ is the exchange-correlation energy per
particle, $V_{ext}$ is the external potential, i.e. the Coulomb
potential of nuclei. To calculate the spin susceptibility we will
assume ${\bf B}_{ext}({\bf r})\rightarrow 0$. It leads to the
effective complete ``non-equilibrium'' field
\begin{equation}
\delta B_{tot}^{\alpha }=\delta B_{ext}^{\alpha }+\frac{\delta B_{xc}^{\alpha }}{%
\delta m^{\beta }}\delta m^{\beta }  \label{linear}
\end{equation}
where $\alpha \beta $ are Cartesian indices and the sum over
repeated indices is assumed.

By definition of the exact non-local frequency-dependent spin
susceptibility $\widehat{\chi }^{\alpha \beta }$ the variation of
the spin density is equal to
\begin{equation}
\delta m^{\alpha }=\widehat{\chi }^{\alpha \beta } \delta
B_{ext}^{\beta } \label{linear1}
\end{equation}
Operator product is defined here as usual:
\begin{equation}
(\widehat{\chi }\varphi )({\bf r)=}\int d{\bf r}^{\prime }\chi ({\bf r,r}%
^{\prime })\varphi ({\bf r}^{\prime })  \label{def}
\end{equation}
On other hand, it was shown in Ref. \onlinecite{TDSDF} that in the
time-dependent density functional theory we should have exactly
\begin{equation}
\delta m^{\alpha }=\widehat{\chi }_{0}^{\alpha \beta } \delta
B_{tot}^{\beta } \label{linear2}
\end{equation}
where $\widehat{\chi }_{0}^{\alpha \beta }$ is the susceptibility
of an auxiliary system of free Kohn-Sham particles. It has been
demonstrated originally for the charge excitations but the
generalization of TDDFT
for the spin-polarized case \cite{vosko}, shows that the equation
(\ref{linear2}) holds also for spin excitations. Comparing two
expressions for $\delta m^{\alpha }$ we have the following equation
\begin{equation}
\widehat{\chi }^{\alpha \beta }=\widehat{\chi }_{0}^{\alpha \beta }+\widehat{%
\chi }_{0}^{\alpha \gamma }\frac{\delta B_{xc}^{\gamma }}{\delta m^{\delta }}%
\widehat{\chi }^{\delta \beta }  \label{dyson}
\end{equation}
This ``RPA-like'' equation is formally exact in ADA-TDDFT. For the
{\it local} spin density approximation (Eq.(\ref{local})) one has
\begin{equation}
\frac{\delta B_{xc}^{\gamma }}{\delta m^{\delta }}=\frac{B_{xc}}{m}\left(
\delta _{\gamma \delta }-\frac{m^{\gamma }m^{\delta }}{m^{2}}\right) +\frac{%
m^{\gamma }m^{\delta }}{m^{2}}\frac{\partial B_{xc}}{\partial m}
\label{local1}
\end{equation}
The first term in Eq.(\ref{local1}) is purely transverse and the
second one is purely longitudinal with respect to the local
magnetization vector. In collinear magnetic structures there are
no coupling between the longitudinal and transverse components and
for the transverse spin susceptibility we have the following
equation:
\begin{equation}
\chi ^{+-}({\bf r,r}^{\prime },\omega )=\chi _{0}^{+-}({\bf r,r}^{\prime
},\omega )+\int d{\bf r}^{\prime \prime }\chi _{0}^{+-}({\bf r,r}^{\prime
\prime },\omega )I_{xc}({\bf r}^{\prime \prime })\chi ^{+-}({\bf r}^{\prime
\prime }{\bf ,r}^{\prime },\omega )  \label{transverse}
\end{equation}
where
\begin{equation}
I_{xc}=\frac{B_{xc}}{m}  \label{stoner}
\end{equation}
is an exchange-correlation ``Hund's rule'' interaction, the magnetic
and charge electron density being defined as usual
\begin{eqnarray}
m=\sum_{\mu \sigma }\sigma f_{\mu \sigma }\mid \psi _{\mu \sigma }({\bf r)}%
\mid ^{2} \nonumber \\
n=\sum_{\mu \sigma } f_{\mu \sigma }\mid \psi _{\mu \sigma }({\bf r)}%
\mid ^{2} \label{magn}
\end{eqnarray}
The bare susceptibility has the following form:
\begin{equation}
\chi _{0}^{+-}({\bf r,r}^{\prime },\omega )=\sum_{\mu \nu
}\frac{f_{\mu \uparrow }-f_{\nu \downarrow }}{\omega-\varepsilon
_{\mu \uparrow }+\varepsilon _{\nu \downarrow }}\psi _{\mu
\uparrow }^{\ast }({\bf r)}\psi _{\nu
\downarrow }({\bf r)}\psi _{\nu \downarrow }^{\ast }({\bf r}^{\prime }{\bf )}%
\psi _{\mu \uparrow }({\bf r}^{\prime }{\bf )}  \label{empty}
\end{equation}
where $\psi _{\mu \sigma }$ and $\varepsilon _{\mu \sigma }$ are eigenstates
and eigenvalues for the Kohn-Sham quasiparticles
\begin{eqnarray}
\left( H_{0} - {\frac{1}{2}} \sigma B_{xc}\right) \psi _{\mu
\sigma }
&=&\varepsilon _{\mu \sigma }\psi _{\mu \sigma }  \nonumber \\
H_{0} &=&-\nabla ^{2}+V({\bf r)}  \label{sham}
\end{eqnarray}
and $f_{\mu \sigma }=f\left( \varepsilon _{\mu \sigma }\right)$ is the
Fermi distribution function.

Although the longitudinal spin susceptibility is not necessary to
consider the exchange interactions it is instructive to write
also an explicit expression for it. The derivation is similar to
those presented above with a small complication, since we have to
consider separately the response of spin-up and spin-down
electrons. Suppose we have an external perturbation $\delta
V_{ext}^{\sigma }$. It leads to the change of the exchange
correlation potential $V_{xc}^{\sigma }=\frac{\partial \left(
n\varepsilon _{xc}\right) }{\partial n_{\sigma }}$ ($n_{\sigma
}=\frac{1}{2}\left( n + \sigma m\right) $), namely,
\begin{eqnarray}
\delta V_{xc}^{\sigma } &=&U_{\sigma \sigma ^{\prime }}\delta
n_{\sigma
^{\prime }},  \label{long} \\
U_{\sigma \sigma ^{\prime }} &=&\frac{\partial ^{2}\left(
n\varepsilon _{xc}\right) }{\partial n_{\sigma }\partial
n_{\sigma ^{\prime }}}  \nonumber
\end{eqnarray}
which gives the total perturbation $\delta V_{tot}^{\sigma
}=\delta V_{ext}^{\sigma }+\delta V_{xc}^{\sigma }$. One can
introduce the response functions $K^{\sigma \sigma ^{\prime }}$ by
the formal expression
\begin{equation}
\delta n^{\sigma }=K^{\sigma \sigma ^{\prime }}\delta V_{ext}^{\sigma
^{\prime }}  \label{K}
\end{equation}
Then the longitudinal spin susceptibility can be expressed in
terms of the $K$-functions as
\begin{equation}
\chi ^{zz}=\frac{1}{4}\left( K^{\uparrow \uparrow }+K^{\downarrow \downarrow
}-K^{\uparrow \downarrow }-K^{\downarrow \uparrow }\right)  \label{long1}
\end{equation}
At the same time, according to the general scheme of
TDDFT, we have
\begin{equation}
\delta n^{\sigma }=\widehat{X}_{\sigma }\delta V_{tot}^{\sigma }
\label{long2}
\end{equation}
where
\begin{equation}
X_{\sigma }\left( {\bf r,r}^{\prime }\right) =%
\sum \limits_{\mu \nu }%
\frac{f_{\mu \sigma }-f_{\nu \sigma }}{\omega-\varepsilon _{_{\mu
\sigma }}+ \varepsilon _{\nu \sigma}}\psi _{_{\mu \sigma }}^{\ast
}({\bf r})\psi _{_{\nu \sigma }}({\bf r})\psi _{_{\mu \sigma
}}({\bf r}^{\prime })\psi _{_{\nu \sigma }}^{\ast }({\bf
r}^{\prime })  \label{empty_long}
\end{equation}
Comparing these two expressions for $\delta n^{\sigma}$ one obtains
\begin{eqnarray}
K^{\uparrow \uparrow } &=&X_{\uparrow }+X_{\uparrow }U_{\uparrow
\uparrow }K^{\uparrow \uparrow }+X_{\uparrow }U_{\uparrow
\downarrow}K^{\downarrow \uparrow }  \nonumber \\
K^{\downarrow \downarrow } &=&X_{\downarrow }+X_{\downarrow
}U_{\downarrow \downarrow }K^{\downarrow \downarrow
}+X_{\downarrow}U_{\downarrow \uparrow }K^{\uparrow \downarrow }  \nonumber \\
K^{\uparrow \downarrow } &=&X_{\uparrow }U_{\uparrow \downarrow
}K^{\downarrow \downarrow }+X_{\uparrow }U_{\uparrow \uparrow
}K^{\uparrow \downarrow } \nonumber \\
K^{\downarrow \uparrow }&=&X_{\downarrow }U_{\downarrow \uparrow
}K^{\uparrow \uparrow }+X_{\downarrow }U_{\downarrow \downarrow
}K^{\downarrow \uparrow } \label{kubo}
\end{eqnarray}
Similar expressions have been obtained in the RPA for the Hubbard
model in Ref \onlinecite{kubo}. A coupling between the longitudinal
spin and density degrees of freedom is important also for the
electronic structure calculations which take into account
correlation effects \cite{KL2002,KL99}.

Let us continue the derivation of useful expression for the
transverse susceptibility (Eq.(\ref{transverse})). In order to
consider the case of small $\omega $ it is useful to make some
identical transformations of the kernel (\ref{empty}) similar to
the Hubbard model consideration \cite{IK1}. Using Eq.(\ref{sham})
one can find
\begin{equation}
B_{xc}\psi _{\mu \uparrow }\psi _{\nu \downarrow }^{\ast }=\left(
\varepsilon _{\nu \downarrow }-\varepsilon _{\mu \uparrow }\right) \psi
_{\nu \downarrow }^{\ast }\psi _{\mu \uparrow }+\nabla (\psi _{\mu \uparrow
}\nabla \psi _{\nu \downarrow }^{\ast }-\psi _{\nu \downarrow }^{\ast
}\nabla \psi _{\mu \uparrow })  \label{iden}
\end{equation}
Substituting Eq.(\ref{iden}) into Eq.(\ref{empty}) we obtain
\begin{equation}
(\chi _{0}^{+-}B_{xc})({\bf r,r}^{\prime },\omega )=m({\bf r)\delta }({\bf %
r-r}^{\prime }{\bf )}-\omega \chi _{0}^{+-}({\bf r,r}^{\prime },\omega )
\label{iden1}
\end{equation}
where we used the completeness condition
\begin{equation}
\sum_{\mu }\psi _{\mu \sigma }^{\ast }({\bf r)}\psi _{\mu \sigma }({\bf r}%
^{\prime }{\bf )=\delta }({\bf r-r}^{\prime }{\bf )}
\label{complet}
\end{equation}
Substituting Eq.(\ref{iden1}) into Eq.(\ref{empty}) we can
transform the latter expression to the following form
\begin{equation}
\widehat{\chi}^{+-}=\widehat{\chi}_{0}^{+-}+\widehat{\chi}_{0}^{+-}\frac{%
B_{xc}}{m}\widehat{\chi}^{+-}=\widehat{\chi}_{0}^{+-}+\widehat{\chi}%
^{+-}-\omega \widehat{\chi}_{0}^{+-}\frac{1}{m}\widehat{\chi}^{+-}+\frac{%
\widehat{\Lambda }}{m}\widehat{\chi}^{+-}  \label{iden2}
\end{equation}
or, equivalently,
\begin{equation}
\widehat{\chi }^{+-}=m\left[\omega -\left( \widehat{\chi _{0}}^{+-}\right) ^{-1}%
\widehat{\Lambda }\right]^{-1}  \label{chi_lambda}
\end{equation}
where
\begin{equation}
\Lambda ({\bf r,r}^{\prime },\omega )=\sum_{\mu \nu }\frac{f_{\mu
\uparrow }-f_{\nu \downarrow }}{\omega-\varepsilon _{\mu \uparrow
}+ \varepsilon _{\nu \downarrow }}\psi _{\mu \uparrow }^{\ast
}({\bf r)}\psi _{\nu \downarrow }({\bf r)}\nabla \left[\psi _{\mu
\uparrow }({\bf r}^{\prime
}{\bf )}\nabla \psi _{\nu \downarrow }^{\ast }({\bf r}^{\prime }%
{\bf )}-\psi _{\nu \downarrow }^{\ast }({\bf r}^{\prime }{\bf
)}\nabla \psi _{\mu \uparrow }({\bf r}^{\prime }{\bf )}\right]
\label{lambda}
\end{equation}
Using Eqs.(\ref{empty}),(\ref{chi_lambda}) one has finally
\begin{equation}
\widehat{\chi }^{+-}=\left( m+\widehat{\Lambda }\right) \left( \omega -I_{xc}%
\widehat{\Lambda }\right) ^{-1}  \label{chi_final}
\end{equation}
which is exactly equivalent to Eq.(\ref{transverse}) but much
more suitable for investigation of the magnon spectrum. Spin wave
excitations can be separated from the Stoner continuum (e.g.,
paramagnons) only in the adiabatic approximation,
which means the replacement $\Lambda ({\bf r,r}^{\prime },\omega )$ by $%
\Lambda ({\bf r,r}^{\prime },0)$ in Eq.(\ref{chi_final}).
Otherwise one should just find the poles of the total
susceptibility, and the whole concept of ``exchange interactions''
is not uniquely defined. Nevertheless, {\it formally} we can
introduce the effective exchange interactions via the quantities
\begin{equation}
\Omega({\bf r,r}^{\prime },\omega )=I_{xc}\Lambda ({\bf
r,r}^{\prime },\omega ). \label{A}
\end{equation}
Substituting Eq.(\ref{iden}) into Eq.(\ref{chi_final}) we get\

\begin{equation}
\Lambda ({\bf r,r}^{\prime },\omega )=\sum_{\mu \nu }\frac{f_{\mu
\uparrow }-f_{\nu \downarrow }}{\omega - \varepsilon _{\mu
\uparrow }+\varepsilon _{\nu \downarrow}}\psi _{\mu \uparrow
}^{\ast }({\bf r)}\psi _{\nu \downarrow }\left[{\bf }B_{xc}({\bf
r}^{\prime })-\varepsilon _{\nu \downarrow}+\varepsilon _{\mu
\uparrow }\right]\psi _{\nu \downarrow }^{\ast }({\bf r}^{\prime
}{\bf )}\psi _{\mu \uparrow }({\bf r}^{\prime }{\bf )}
\label{lambda1}
\end{equation}
Therefore
\begin{equation}
\Omega({\bf r,r}^{\prime },\omega )=\frac{4}{m({\bf r)}}J({\bf
r,r}^{\prime },\omega ) + I_{xc}({\bf r)}\sum_{\mu \nu
}\frac{f_{\mu \uparrow }-f_{\nu \downarrow }}{\omega-\varepsilon
_{\mu \uparrow }+\varepsilon _{\nu \downarrow }}\left(
\varepsilon _{\mu \uparrow}-\varepsilon _{\nu \downarrow }\right)
\psi _{\mu \uparrow }^{\ast }({\bf r)}\psi _{\nu \downarrow
}({\bf r)}\psi _{\nu \downarrow }^{\ast }({\bf r}^{\prime }{\bf
)}\psi _{\mu \uparrow }({\bf r}^{\prime }{\bf )}  \label{A1}
\end{equation}
where an expression for frequency dependent exchange interactions
has the following form
\begin{equation}
J({\bf r,r}^{\prime },\omega )=\frac{1}{4}\sum_{\mu \nu
}\frac{f_{\mu \uparrow }-f_{\nu \downarrow }}{\omega-\varepsilon
_{\mu \uparrow }+ \varepsilon _{\nu \downarrow}}\psi _{\mu
\uparrow }^{\ast }({\bf r)}B_{xc}({\bf r)}\psi _{\nu \downarrow
}({\bf r)}\psi _{\nu \downarrow }^{\ast }({\bf r}^{\prime }{\bf
)}B_{xc}({\bf r}^{\prime }{\bf )}\psi _{\mu \uparrow }({\bf
r}^{\prime }{\bf )}  \label{J}
\end{equation}
The later coincides with the exchange integrals
\cite{LIXT,JMMM,Licht95} if we neglect the $\omega $ -
dependence. Since $B_{xc}\sim m$ we have $J \sim m^2$ and the
expression (\ref{A1}) vanishes in non-magnetic case, as it should be.
Using the identity (\ref{complet}) one can show that
\begin{equation}
\Omega({\bf r,r}^{\prime },0)=\frac{4}{m({\bf r)}}J({\bf r,r}^{\prime },0)-B_{xc}
({\bf r)\delta }({\bf r-r}^{\prime })  \label{A2}
\end{equation}
Note that for \ $\omega =0$ \ we have exactly:
\begin{equation*}
-I_{xc}\widehat{\Lambda }=\widehat{\Omega}
\end{equation*}
and the static susceptibility $\widehat{\chi }^{+-}\left( 0\right)$ can be
represented in the form
\begin{equation}
\widehat{\chi }^{+-}\left( 0\right) =m\left(
\widehat{\Omega}^{-1}-{B_{xc}^{-1}}\right)  \label{static}
\end{equation} which is equivalent to the
result of Ref. \onlinecite{bruno}
\begin{equation}
\widehat{\widetilde{\Omega}}=\widehat{\Omega}\left(
1-B_{xc}^{-1}\widehat{\Omega}\right) ^{-1}  \label{bruno}
\end{equation}
for the renormalized exchange interaction if one define them in
terms of inverse {\it static} susceptibility \cite{antr,logan}.

As it was stressed above for a generic case of an itinerant
electron magnet it is impossible to introduce the effective
exchange integrals and one should to work with the generalized
spin susceptibility. Any definition of the exchange integrals
assume the adiabatic approximation somewhere. For the spin-wave
spectrum which is determined by the pole of the transverse
susceptibility it is natural to formulate ``exchange concept'' as
neglecting of the $\omega $-dependence in $\widehat{\Omega}$.
Then, in virtue of Eq.(\ref{chi_final}) the magnon frequencies
are just eigenstates of the operator $\widehat{\Omega}\left(
0\right) $ which exactly corresponds to the expression from the
``old'' MFT exchange interactions \cite{LIXT,JMMM}. Note that for
the long-wavelength limit ${\bf q}\rightarrow 0$ this result turns out
to be exact which proves the above statement about the stiffness
constant $D$: in the framework of the local approximation it is
rigorous. Corrections to $D$ from a {\it nonlocality} of the
exchange-correlation potential have been estimated recently
\cite{KA} for Fe and Ni and turned out to be small.

At the same time, if we are interested in the computations of the
thermodynamic properties such as the Curie temperature $T_{C}$ the
renormalized exchange integrals can really give more accurate
results. One can introduce for the itinerant electron magnets a
magnon-like operators
\begin{equation}
b_{{\bf q}}=\frac{1}{\sqrt{m_{0}}}S_{{\bf q}}^{-}, \ \ \ b_{{\bf q}}^{\dagger }=%
\frac{1}{\sqrt{m_{0}}}S_{{\bf q}}^{+}  \label{holstein}
\end{equation}
where $m_{0}=2\overline{S}$ is the ground-state magnetization and write for
the temperature dependence of the magnetization the Bloch-like expression
\begin{equation}
m\left( T\right) =m_{0}-\sum\limits_{{\bf q}}\left\langle b_{{\bf q}%
}^{\dagger }b_{{\bf q}}\right\rangle =m_{0}+\frac{1}{m_{0}}\sum\limits_{{\bf %
q}}\int\limits_{-\infty }^{\infty }d\omega \frac{%
%TCIMACRO{\func{Im}}%
%BeginExpansion
\mathop{\rm Im}%
%EndExpansion
\chi ^{+-}({\bf q,}\omega )}{\exp \left( \omega /T\right) -1}  \label{Bloch}
\end{equation}
(a similar approximation for the Hubbard model has been proposed
in Ref. \onlinecite{IK1}). If we will use the classical-spin
approximation usually exploited for first-principle estimations
of the Curie temperature one should replace the Planck function
in Eq.(\ref{Bloch}) by its classical limit $T/\omega $ which
immediately gives (taking into account the Kramers-Kronig relations)
the following expression for the Curie
temperature
\begin{equation}
\frac{1}{T_{C}}=\frac{1}{m_{0}^{2}}\sum\limits_{{\bf q}}\chi ^{+-}({\bf
q,}\omega =0) \label{T_C}
\end{equation}
which is identical to the expression from Ref. \onlinecite{bruno} in
terms of the {\it renormalized} exchange interactions. Note,
however, that the quantum character of the spin (which can be
taken into account only beyond the LSDA) is probably very
essential for proper description of high-temperature magnetism of
transition metals \cite{ourFeNi} which makes a problem of an
improvement of {\it classical} estimations of $T_{C}$ less
important.

In order to test different approximations to the exchange interactions we
calculated spin-wave spectrum for iron and nickel using the LMTO-TB
method \cite{LMTO}. The orthogonal LMTO representation was used and
the calculation scheme was the following.

The matrix of Green function in s,p,d-basis set is equal to

\begin{equation}
G_{\sigma }({\bf k},\omega _{n})=\left[ i\omega _{n}+\mu -H_{LDA}^{\sigma }(%
{\bf k})\right] ^{-1}  \label{green}
\end{equation}
where $\mu $ is the chemical potential, $\omega _{n}$ are Matsubara
frequencies and $H_{LDA}$ is the orthogonal LSDA Hamiltonian. We use the
following approximation for the matrix $I_{xc}$:

\begin{equation}
V_{xc}\equiv mI_{xc}=H_{LDA}^{\uparrow }(0)-H_{LDA}^{\downarrow }(0)
\label{splitting}
\end{equation}
The matrix of the LDA-susceptibility has been calculated using the fast
Fourier transform technique with $k=({\bf k},\omega _{n})$

\begin{equation}
\chi _{0}^{+-}(q)=-\sum_{k}G_{\uparrow }(k)\ast G_{\downarrow
}(k+q) \label{chizero}
\end{equation}
The Fourier transforms of the ``bare'' exchange interactions
$\widehat{J}\left( 0\right) $ (\ref{J}) is defined as

\begin{equation}
J({\bf q})=\frac{1}{4}V_{xc}\chi _{0}^{+-}({\bf q,}\omega =0)V_{xc}
\label{J1}
\end{equation}
whereas for the ``renormalized'' exchange integrals \cite{bruno,antr} one
has
\begin{eqnarray}
\widetilde{J}(0)-\widetilde{J}({\bf q})&=&\frac{1}{4}Tr_L \left[ m \left( \chi ^{+-}(%
{\bf q,}\omega =0)\right) ^{-1}m \right] \nonumber \\
&=&\frac{1}{4}Tr_L \left[ m\left( \left( \chi _{0}^{+-}(%
{\bf q,}\omega =0)\right) ^{-1}-\left( \chi _{0}^{+-}({\bf
q}=0{\bf ,}\omega =0)\right) ^{-1}\right) m \right]
\label{renormJ}
\end{eqnarray}
The magnon spectrum is determined via the exchange integrals as
\begin{equation}
\omega ({\bf q})=\frac{4}{M}\left[ J(0)-J({\bf q})\right] ,
\label{omega}
\end{equation}
where $M=Tr_L m$ is the total magnetic moment.
 One can see from Fig.1 and Fig.2 that the LDA ``bare'' exchange
parameter better describe the spin-wave spectrum in Fe and Ni,
while thermodynamics (e.g. the Curie temperature - see Table. I)
are more reasonable with the exact static LDA-exchange (which is
the ``RPA''-like expression). This confirms a general
consideration presented above.

%%%%% Fig.1 %%%%%%%%%%%%%%%%%%%%%%%%%%%%%%%%%%%%%
\begin{figure}[tbp]
\includegraphics[width=.6\textwidth]{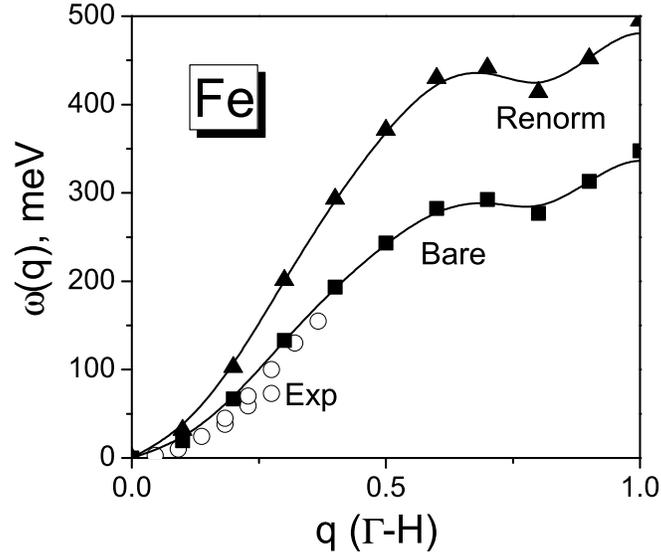}
\caption{ The spin-wave spectrum for ferromagnetic iron in the
bare exchange (Eq.(\ref{J1})) and renormalized exchange
(Eq.(\ref{renormJ})) scheme in comparison with experimental data
(from Ref.\cite{exp}). } \label{swFe}
\end{figure}
%%%%%%%%%%%%%%%%%%%%%%%%%%%%%%%%%%%%%%%%%
%%%%% Fig.2 %%%%%%%%%%%%%%%%%%%%%%%%%%%%%%%%%%%%%
\begin{figure}[tbp]
\includegraphics[width=.6\textwidth]{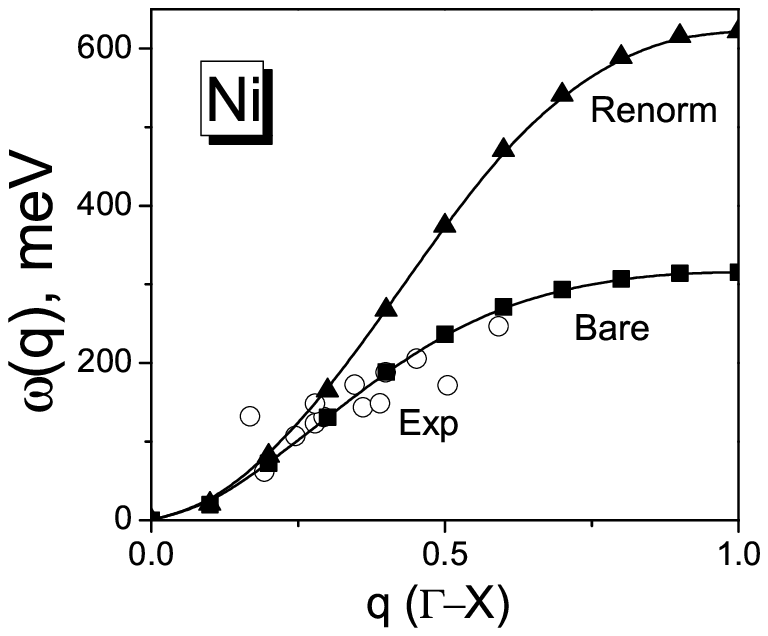}
\caption{ The spin-wave spectrum for ferromagnetic nickel in the
bare exchange (Eq.(\ref{J1})) and renormalized exchange
(Eq.(\ref{renormJ})) scheme in comparison with experimental data
(from Ref.\cite{exp}). } \label{swNi}
\end{figure}
%%%%%%%%%%%%%%%%%%%%%%%%%%%%%%%%%%%%%%%%%

\begin{table}[tbp]
\caption{ Curie temperature (in K) for Fe and Ni calculated with
the bare (Eq.(\ref{J1})) and renormalised (Eq.(\ref{renormJ}))
LDA exchange interactions; ``M'' is the mean-filed approximation
$T_{c}=2/3J(0)$ and ``T'' is the Tjablikov, or RPA, approximation
for $T_{c}$, similar to Eq.(\ref {T_C})} \label{JJ}\vskip 1.0cm
\begin{tabular}{|c|r|r|r|r|r|}
\hline $T_c(K)$ & Exp & Bare-M & Renorm-M & Bare-T & Renorm-T \\
\hline Fe & 1045 & 1060 & 1620 & 820 & 1280 \\ \hline Ni & 631 &
310 & 760 & 285 & 630 \\ \hline
\end{tabular}
\end{table}

\end{document}